\documentclass[aps,prl,twocolumn,superscriptaddress,floatfix]{revtex4}
\usepackage{graphicx,epsfig}
\usepackage{amssymb,times}
\usepackage{amsmath,amsfonts,bm}
\usepackage{color}
\usepackage{epstopdf}

\newcommand{\comment}[1]{{\bf #1}}

\def\be{\begin{equation}}
\def\ee{\end{equation}}

\def \rev@citealpnum#1{{\onlinecite{#1}}}
\def \yd{{^\dagger}}

\def \comment#1{{}}

\def \subsection#1{{\bf #1.}}

\def \atanu#1{{\bf #1}}

\begin{document}
	
\title{\atanu{Characterizations of prethermal states in  periodically driven many-body systems with unbounded chaotic diffusion}}
\author{Atanu Rajak}
\affiliation{Department of Physics, Bar-Ilan University, Ramat Gan 5290002, Israel}
\affiliation{Center for Quantum Entanglement Science and Technology, Bar-Ilan University, Ramat Gan 5290002, Israel}
\affiliation{Presidency University, Kolkata, West Bengal 700073, India}
\author{Itzhack Dana}
\affiliation{Department of Physics, Bar-Ilan University, Ramat Gan 5290002, Israel}
\author{Emanuele G. Dalla Torre}
\affiliation{Department of Physics, Bar-Ilan University, Ramat Gan 5290002, Israel}
\affiliation{Center for Quantum Entanglement Science and Technology, Bar-Ilan University, Ramat Gan 5290002, Israel}

\begin{abstract}

We introduce well-defined characterizations of prethermal states in realistic periodically driven many-body systems with unbounded chaotic diffusion of the kinetic energy. These systems, interacting arrays of periodically kicked rotors, are paradigmatic models of many-body chaos theory. We show that the prethermal states in these systems are well described by a generalized Gibbs ensemble based essentially on the average Hamiltonian. The latter is the quasi-conserved quantity in the prethermal state and the ensemble is characterized by the temperature of the state. An explicit exact expression for this temperature is derived. Using also arguments based on chaos theory, we demonstrate that the lifetime of the prethermal state is exponentially long in the inverse of the temperature, in units of the driving frequency squared. Our analytical results, in particular those for the temperature and the lifetime of the prethermal state, agree well with numerical observations. 
\end{abstract}

\maketitle

When a periodic drive acts on a closed many-body system, it generically leads to heating. The heating hinders the applicability of periodic drives to  tune closed many-body systems and to create new phases of matter, such as time crystals, Floquet topological phases, and other. Earlier studies were able to determine specific situations where the heating can be reduced or suppressed: (i) integrable models, where the heating is restricted by the conserved quantities of the dynamics (see Ref.~\cite{gritsev2017integrable} and references therein); (ii) many-body localized systems, where the disorder prevents the entropy growth associated with heating \cite{ponte15many,lazarides15fate,ponte15periodically,abanin16theory,agarwal17localization,dumitrescu2017logarithmically}; and (iii) high frequency drives \cite{choudhury14stability,bukov15prethermal,citro2015dynamical,goldman15periodically,chandran16interaction,
lellouch17parametric,lellouch2018parametric,abanin15exponentially,kuwahara16floquet,mori2016rigorous,abanin17rigorous,
rajak2018stability,howell2019asymptotic,mori2018floquet}. 

Concerning the last scenario, several studies considered quantum systems with a {\it finite} frequency bandwidth $\Lambda$. It was observed that parametric drives can lead to dynamical instabilities only if the driving frequency $\Omega$ is smaller than $2\Lambda$. According to this criterion, many-body systems are expected to be linearly stable and avoid heating when $\Omega>2\Lambda$ \cite{choudhury14stability,bukov15prethermal,citro2015dynamical,goldman15periodically,chandran16interaction,
lellouch17parametric,lellouch2018parametric}. Recent studies aimed to extend this approach to non-linear effects. In particular, Refs. \cite{abanin15exponentially,kuwahara16floquet,mori2016rigorous,abanin17rigorous} demonstrated rigorously that at large driving frequencies the heating rate is suppressed as $\exp(-\Omega/\Lambda)$. This leads to prethermal states, i.e., almost no heat absorbtion on long time intervals, reflecting the quasi-conservation of the Floquet Hamiltonian. The derivation of the exponential bound above is based on the observation that in order to absorb a quantum from the pump, whose energy is $\hbar \Omega$, it is necessary to consider a perturbation of order $n \approx \hbar\Omega/\hbar \Lambda$. According to time-dependent perturbation theory, the rate of this process is proportional to $\epsilon^{n}$, where $\epsilon$ is determined by the pump intensity. For weak pumps, this leads to the exponential bound above. Interestingly, although quantum mechanics was used in the derivation \cite{note}, the final result does not depend on $\hbar$. This cancellation suggested that the exponential bound should be valid for classical systems as well~\cite{rajak2018stability,howell2019asymptotic,mori2018floquet,note}.

The above-mentioned rigorous bounds have a major limitation: They rely on the assumption of a finite bandwidth and, hence, are valid only for systems whose energy density (energy per particle) is bounded. This assumption holds for common systems such as spin models or non-interacting band models but does not generically apply to realistic many-body systems. In the latter systems, the energy density is not bounded from above either due to a realistic kinetic-energy term or due to the possibility of increasing the interaction potential energy by enlarging the local density of particles.

\begin{figure}[b]
	\includegraphics[scale=0.3]{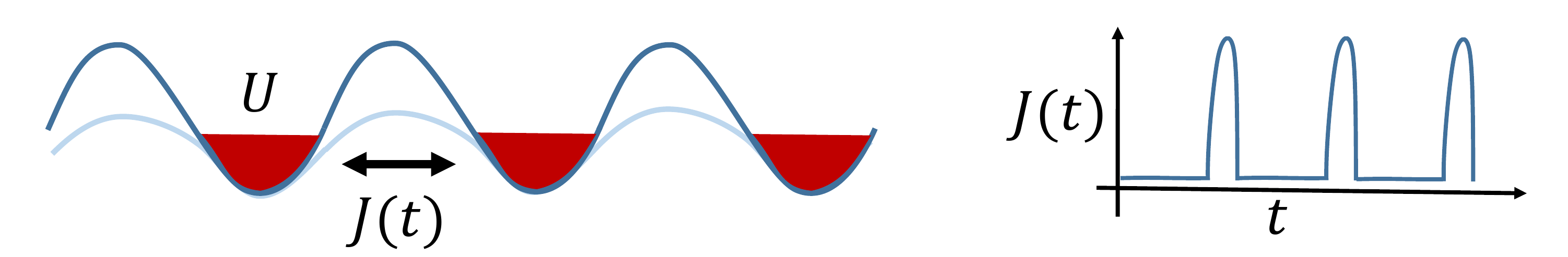}
	\caption{Proposed realization of the system (\ref{eq:H}) using bosonic Josephson junctions created by Bose-Einstein condensates in optical lattices \cite{cataliotti2001josephson}. See note \cite{notee} for details.}
	\label{fig:lattice}
\end{figure}

In this paper, we introduce well-defined characterizations of prethermal states in realistic periodically driven systems with infinite energy density, reflected in an unbounded chaotic diffusion of the kinetic energy at very long times. These are classical systems described by the Hamiltonian
\be
H(t)=\sum_{j=1}^N \left[ \frac{p_j^2}{2} -\kappa \Delta(t) \cos(\phi_j-\phi_{j+1})\right] \label{eq:H}\;.
\ee
Here $p_j$ and $\phi_j$, $j=1,...,N$, are the angular momenta and angles of $N$ rotors, $\kappa$ is a parameter, and $\Delta(t)=\sum_n \delta(t-n\tau)$ is a periodic delta function with period $\tau$. The angles $\phi_j$ satisfy periodic boundary conditions, $\phi_{N+1}=\phi_1$. Importantly, since the kinetic-energy term in Eq. (\ref{eq:H}) is not bounded, one 
can have an infinite energy density. The systems (\ref{eq:H}) are paradigmatic models of many-body chaos theory, introduced in Refs. \cite{kaneko89diffusion,konishi90diffusion} and studied in Refs. \cite{kaneko89diffusion,konishi90diffusion,falcioni91ergodic,chirikov1993theory,chirikov97arnold,mulansky11strong}; see also below. These systems can be  experimentally realized using an array of bosonic Josephson junctions \cite{cataliotti2001josephson}, see Fig.~\ref{fig:lattice} and note \cite{notee}.

The existence of an unbounded chaotic diffusion of the kinetic energy at very long times was established numerically in Refs. \cite{kaneko89diffusion,konishi90diffusion,falcioni91ergodic,chirikov1993theory,chirikov97arnold,mulansky11strong}. This is an Arnol'd diffusion \cite{via}, occurring for arbitrarily small $\kappa$ already for $N>2$ and reflecting the infinite energy density of the system. Despite this, it was shown numerically in work \cite{rajak2018stability} that, at least for some specific choices of initial conditions, the unbounded-diffusion regime was preceded by a long-lived prethermal state in which the kinetic energy almost does not change, see Fig. 2; the lifetime of this state increases exponentially in $1/K$, where $K=\kappa\tau$ is the dimensionless nonintegrability parameter, the only parameter appearing in the Poincar\'{e} map of the system \cite{rajak2018stability}. However, the reason and physical origin of this effect were not clarified.

We consider here in detail the nature of the prethermal states in the systems (\ref{eq:H}) and explain why the inhibition of heating in these states occurs for exponentially long times in $1/K$. Our approach is based on the assumption that a prethermal state, if it exists, can be described by a canonical ensemble like equilibrium (time-independent) systems \cite{lprv,jea}. This is a generalized Gibbs ensemble (GGE) \cite{rigol2007relaxation}, based on constants of the motion, mainly the average Hamiltonian of the system. The GGE is characterized by the temperature $T^*$ of the prethermal state. We derive an explicit exact expression for the statistical quantity $T^*$ in terms of the fully deterministic system parameters. Then, using the GGE and arguments based on many-body chaos theory, we demonstrate that the lifetime of the prethermal state grows exponentially in $1/K$. For times larger than this lifetime, the system exhibits unbounded diffusion by absorbing energy at almost a constant rate. Our analytical results for the GGE, $T^*$, and the lifetime agree well with numerical observations.

\begin{figure}[t]
\includegraphics[scale=0.55]{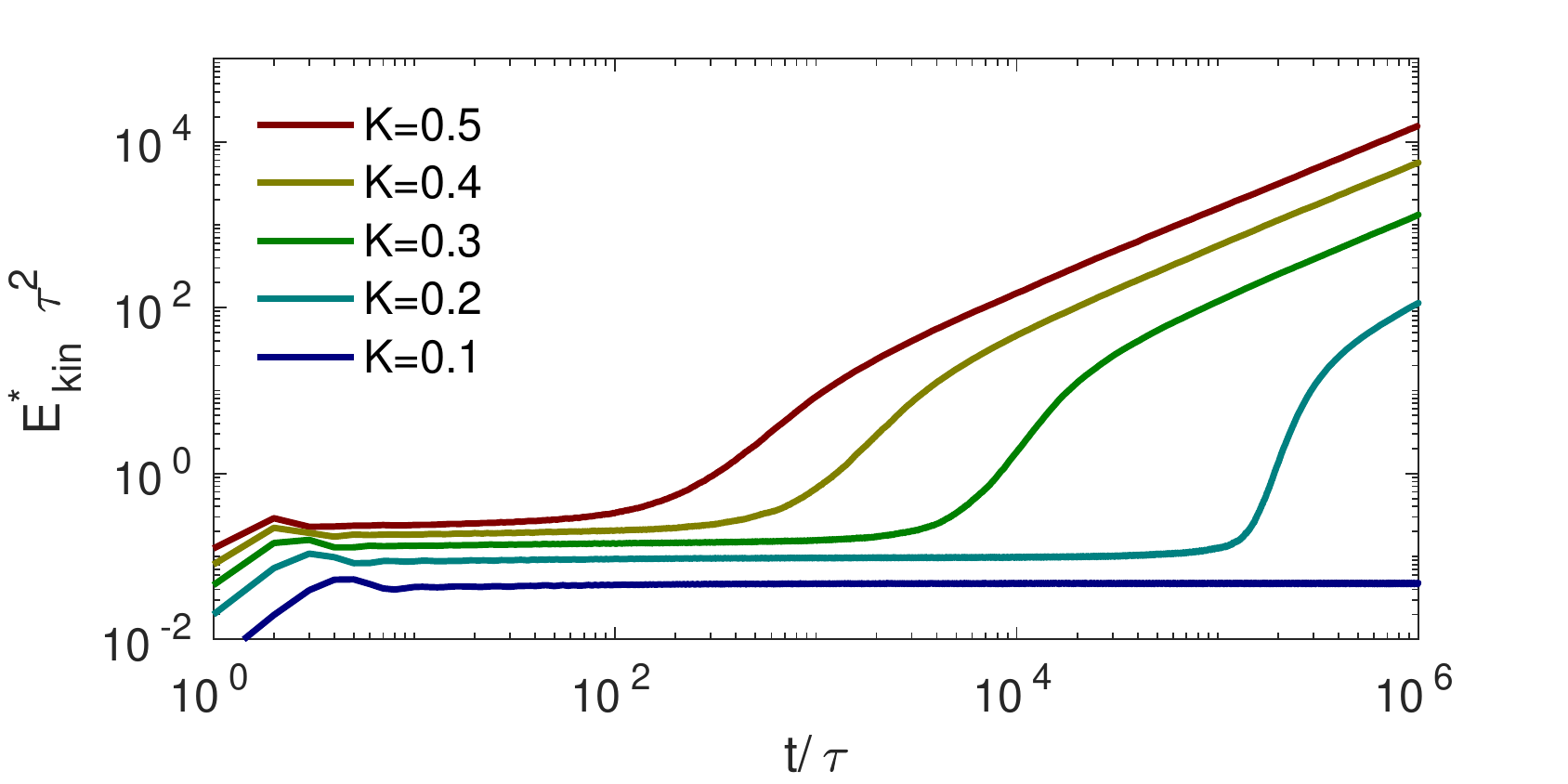}
\caption{Time evolution of the average kinetic energy per rotor, $E_{\rm kin}(t)=(1/N)\sum_{j=1}^N\langle p_j^2(t)/2\rangle$, $N=400$, where the average is over initial conditions with $p_j(t=0)=0$ and $100$ values of $\phi_j(t=0)$ homogeneously distributed between $0$ and $2\pi$. The short-time dynamics is followed by a prethermal plateau, whose lifetime grows exponentially with $1/K$, $K=\kappa\tau$. The plots in this figure appear to be almost insensitive to the number $N$ of rotors, for sufficiently large $N$.}
\label{fig:time_1d}
\end{figure} 

\begin{figure}[t]
	\includegraphics[scale=0.5]{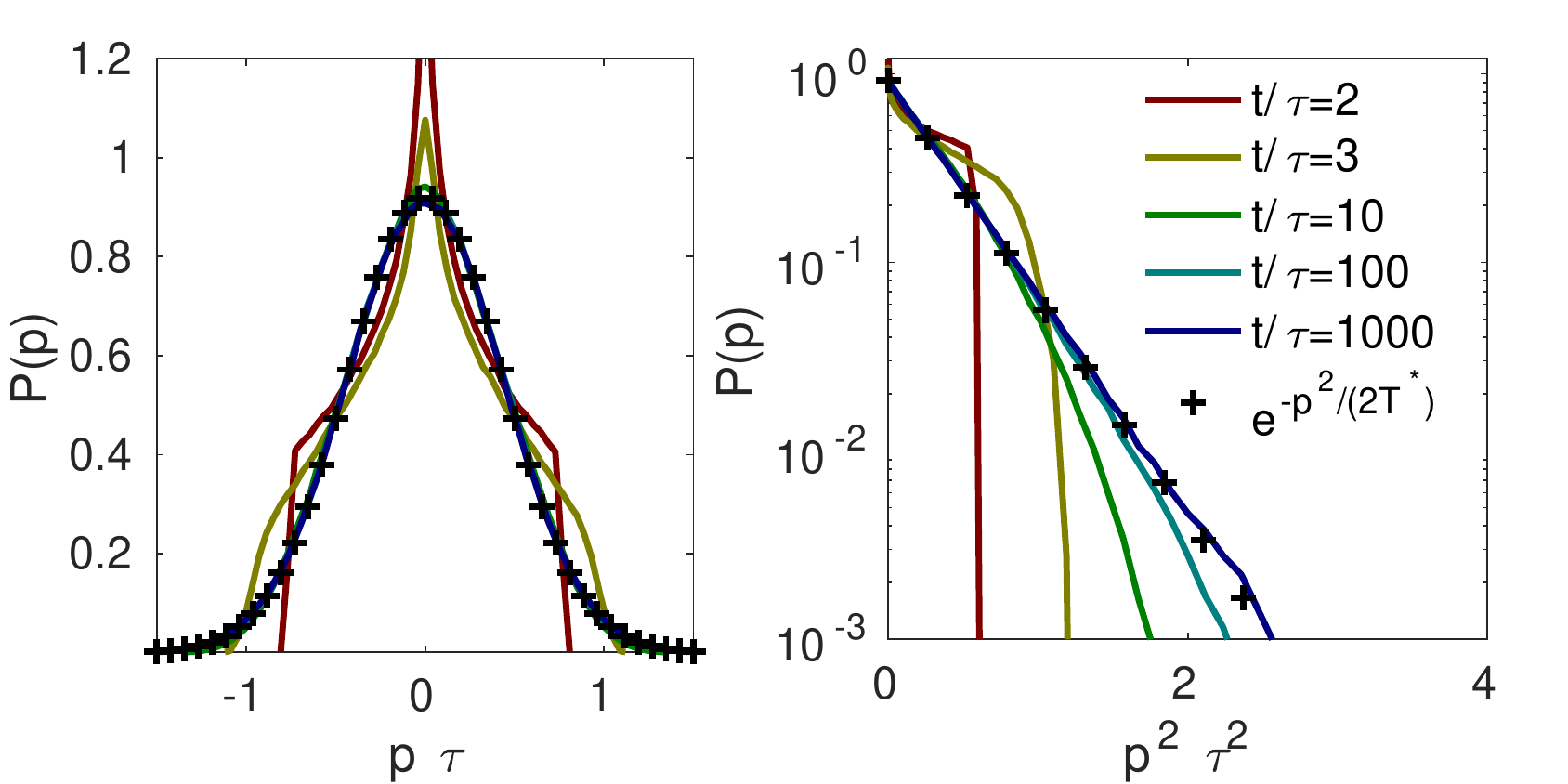}
	\caption{Time evolution of the distribution function of $p_j(t)$ for $N=400$ and $K=0.2$, starting from initial conditions with $p_j(t=0)=0$ and ~$5000$ values of $\phi_j(t=0)$ homogeneously distributed between $0$ and $2\pi$. In the prethermal state, the distribution is Gaussian ('+' points).}
	\label{fig:distr}
\end{figure}

In order to characterize a prethermal state by a quasi-equilibrium GGE, let us identify the constants or quasi-constants of motion in this state. First, a quasi-conserved quantity in the large-frequency regime of $\Omega =2\pi /\tau\gg 1$ is the Floquet Hamiltonian, whose lowest order term is the average Hamiltonian:
\be \label{eq:Hstar} 
H^*=\frac{1}{\tau} \int_0^{\tau} H(t)dt = \sum_{j=1}^N \left[ \frac{p_j^2}{2} - \frac{\kappa}{\tau} \cos(\phi_j-\phi_{j+1})\right].
\ee   
Second, since both Hamiltonians (\ref{eq:H}) and (\ref{eq:Hstar}) depend on the angles $\phi_j$ only via the differences $\phi_j-\phi_{j+1}$, they are invariant under the global translation $\phi_j\rightarrow \phi_j+\chi$, for arbitrary $\chi$. This implies the existence of an exact constant of the motion [besides the approximate one (\ref{eq:Hstar})], i.e., the angular momentum of the center of mass:
\be \label{pcm}
\bar{p}=\frac{1}{N}\sum_{j=1}^N p_j .
\ee
Because of the constant of the motion (\ref{pcm}), the Hamiltonian (\ref{eq:Hstar}) is nonintegrable and completely chaotic for $N>3$ and arbitrarily small $\kappa$, due to Arnol'd diffusion. Thus, for sufficiently large $N$, the prethermal state associated with this Hamiltonian should be well described statistically. This description is given by a GGE defined by a probability distribution featuring the above constants of motion:
\be \label{Pstar}
P^*(\{p_j,\phi_j\})=\frac{1}{Z} \exp\left[ -\frac{H^*(\{p_j,\phi_j\})}{T^*}+\gamma \bar{p}(\{p_j\})\right ],
\ee
where $Z$ is a normalization constant (the partition function),  $T^*$ is the temperature of the prethermal state, and $\gamma$ is some constant (we work in units such that the Boltzmann constant is $k_{\rm B}=1$). Using Eqs. (\ref{eq:Hstar}) and (\ref{pcm}), one can write a more compact expression for the distribution (\ref{Pstar}):
\be \label{Pstarn}
P^*(\{p_j,\phi_j\})=\frac{1}{Z} \exp\left[ -\frac{H^*(\{p_j-\tilde{p},\phi_j\})}{T^*}\right ],
\ee
where $\tilde{p}=\gamma T^*/N$. Since $H^*$ in Eq. (\ref{eq:Hstar}) is the sum of two terms that depend on $p_j$ and 
$\phi_j$ separately, these two sets of variables are statistically independent \cite{note1}. Hence, a probability distribution of angular momenta is well defined:
\be \label{eq:Pstar1}
P^* (\{p_j \}) = Z^{-1} \prod_{j=1}^N \exp \left[ -\frac{(p_j-\tilde{p})^2}{2T^*}\right] . 
\ee
Thus, starting from an ensemble of initial conditions all with $p_j=\tilde{p}$, $j=1,...,N$, the final distribution of $p_j$ in the prethermal state should be a Gaussian centered on $p_j=\tilde{p}$. This is fully confirmed by numerical experiments, see Fig. \ref{fig:distr} for $\tilde{p}=0$. This figure also shows that at short times, before the prethermal state, the function $P^* (\{p_j \})$ is non-universal and depends on time. One can exactly calculate from Eq. (\ref{eq:Pstar1}) the average kinetic energy per rotor in the prethermal state:
\be \label{Ekin}
E_{\rm kin}=\frac{1}{N}\sum_{j=1}^N\int \frac{p_j^2}{2Z}\prod_{j'=1}^N e^{-(p_{j'}-\tilde{p})^2/(2T^*)}dp_{j'}=\frac{T^*+\tilde{p}^2}{2}.
\ee
See also note \cite{note2}.

\begin{figure}[t]
	\includegraphics[scale=0.5]{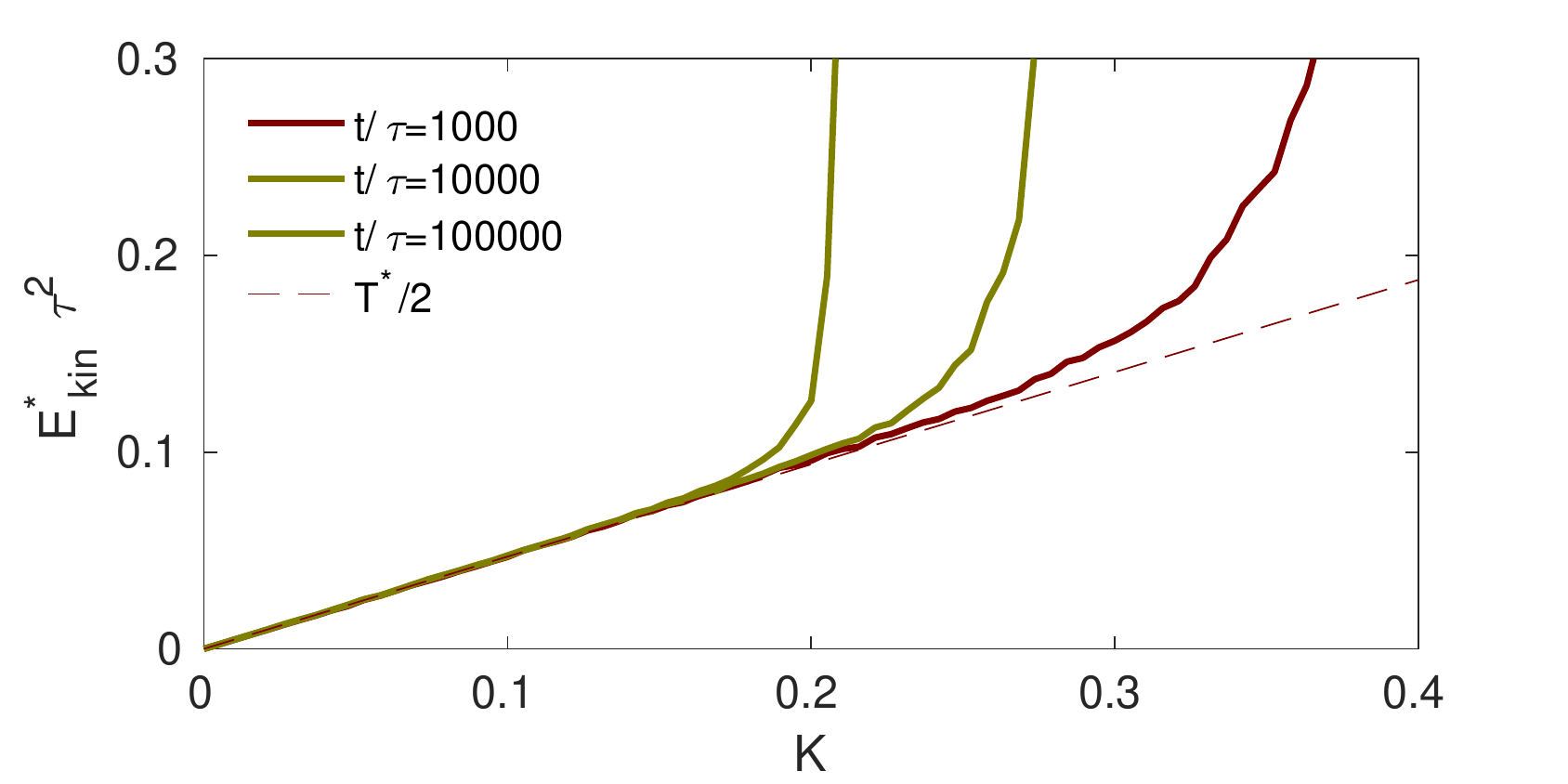}
	\caption{Average kinetic energy per rotor (in unit of $\tau^2$) as a function of $K$ for $\tilde{p}=0$ and three values 
		of $t/\tau$. Dashed line: theoretical prediction, $E_{\rm kin}^*\tau^2=\tau^2 T^*/2=0.469K$, where $T^*$ is given by Eq. (\ref{Tstar}). A good agreement with the theoretical prediction is observed in the region of $(K,t/\tau )$ values corresponding to prethermal states. This region extends to larger values of $t/\tau$ as $K$ decreases.}
	\label{fig:Estar_1d}
\end{figure}

The temperature of the prethermal state $T^*$ will be determined by the quasi-conservation of the time-independent average Hamiltonian $H^*$. As we now show, $T^*$ can be computed analytically by equating the energy of the initial state, $E_0$, with the average energy of the prethermal state, $E^*$. The latter is given by the sum of two terms [see 
Eq.~(\ref{eq:Hstar})]:
\be \label{eq:Estar}
E^* = \sum_{j=1}^N\frac{\left\langle {p^2_j}\right\rangle_*}2 -\frac{\kappa}{\tau}\sum_{j=1}^N\left\langle\cos(\phi_j-\phi_{j+1})\right\rangle_* \ . 
\ee
Here $\langle O \rangle_* = \int \prod_{j=1}^N O(\{p_j,\phi_j\})~P^*(\{p_j,\phi_j\})~dp_j~d\phi_j$ is the average over the prethermal state. The first term on the right-hand-side of Eq. (\ref{eq:Estar}) is equal to $NE_{\rm kin}$, where $E_{\rm kin}$ is given by Eq. (\ref{Ekin}). To evaluate the second term, we transform to the relative coordinates $\varphi_j = \phi_j - \phi_{j+1}$ (the constant Jacobian can be absorbed in the partition function), define $\epsilon=\kappa/(\tau T^*)$, and use
\begin{align}\label{ident}
\langle\cos(\varphi_j)\rangle_* & = \frac{\int_0^{2\pi} d\varphi_j \cos(\varphi_j) e^{\epsilon\cos\varphi_j}}{\int_0^{2\pi} d\varphi_j e^{\epsilon\cos\varphi_j}}
 = \frac{I_1(\epsilon)}{I_0(\epsilon)},
\end{align}
where $I_n(x)$ is the modified Bessel function of order $n$. We then get \cite{note3}: 
\be \label{eq:Estarn}
E^*  = N\frac{T^*+\tilde{p}^2}{2}-\frac{\kappa}{\tau}\frac{I_1(\epsilon)}{I_0(\epsilon)}\ .
\ee
The initial conditions have $p_j=\tilde{p}$ and $\phi_j$ homogeneously distributed from $0$ to $2\pi$, $j=1,...,N$. Therefore, the energy of the initial state is simply $E_0=N\tilde{p}^2/2$. By equating $E_0$ with $E^*$ in Eq. (\ref{eq:Estarn}), we obtain the equation $2\epsilon I_1(\epsilon)= I_0(\epsilon)$ for $\epsilon$. The numerical solution of this equation is $\epsilon=1.066$ or, since $\epsilon=\kappa/(\tau T^*)$,
\be \label{Tstar}
T^*=\frac{\kappa}{1.066\tau}=0.9381\frac{K}{\tau^2}\ ,
\ee
where $K=\kappa\tau$. The main result (\ref{Tstar}) is numerically confirmed by Fig. 4, showing $E_{\rm kin}\tau^2$ versus $K$ for $\tilde{p}=0$ [then $E_{\rm kin}=T^*/2$ by Eq. (\ref{Ekin})].

\begin{figure}[t]
	\includegraphics[scale=0.5]{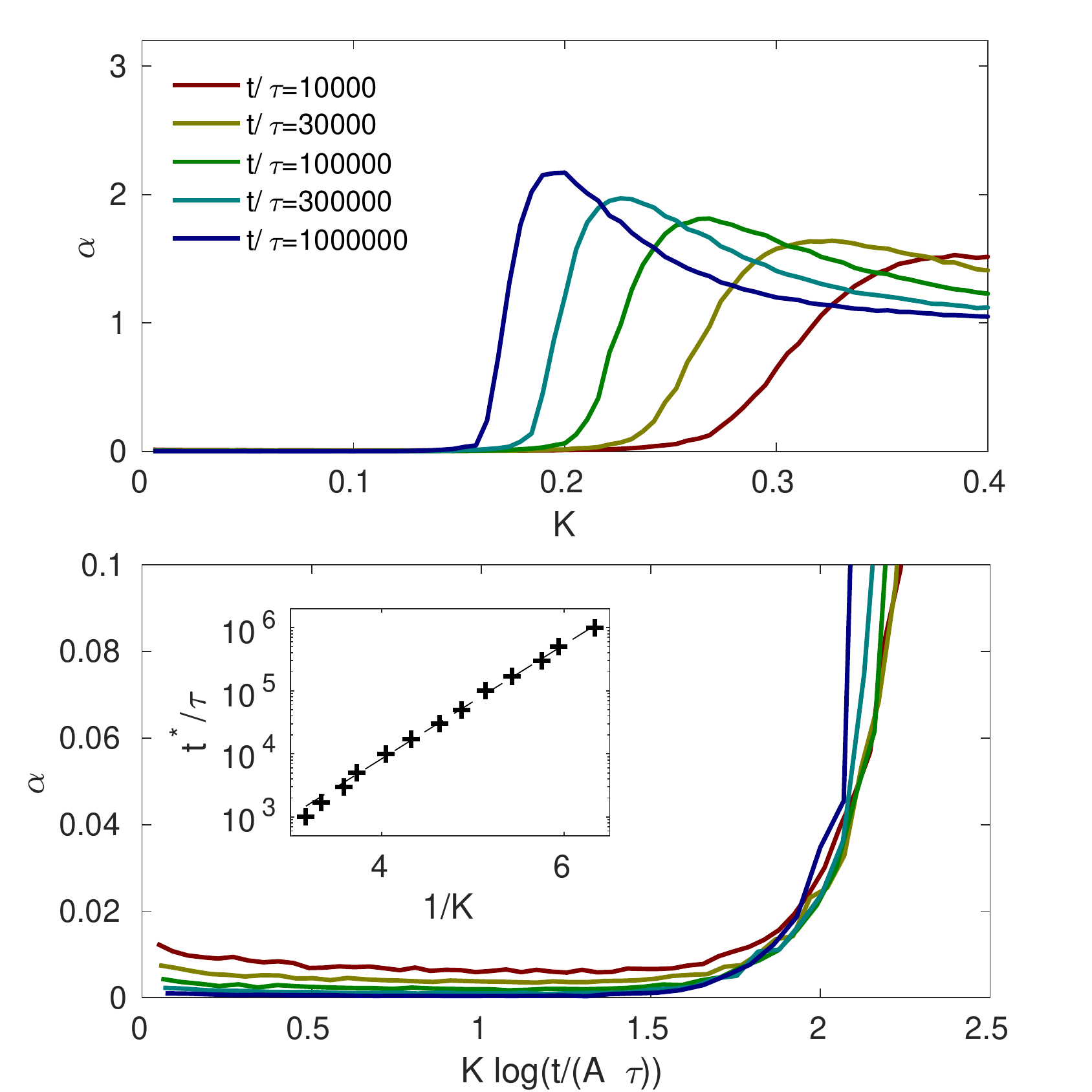}
	\caption{Upper panel: Diffusion exponent $\alpha$ [obtained by fitting the function $E_{\rm kin}(t)$ to $C t^{\alpha}$] plotted versus $K$ for different values of $t/\tau$. Lower panel: The small-$K$ behaviors collapse when $\alpha$ is plotted versus $K{\rm log}(t/(A\tau))$. Here $A$ is determined from the best fit of the lifetimes $t^*/\tau$, at which $\alpha=0.05$, as a function of $K$; see inset. The dashed line in the inset is the exponential fit $t^*/\tau= A\exp(B/K)$ with $A=2.04$ and $B=2.08$.}
	\label{fig:pretherm_1d}
\end{figure}

Having obtained a full characterization of the prethermal state, we now address its stability to the periodic drive. To get a quantitative measure of the lifetime of the prethermal state, we fit the time evolution of $E_{\rm kin}$ up to time $t$ by a power law $t^\alpha$ and monitor $\alpha$ as a function of $K$ for different values of $t/\tau$. The resulting plots (curves) are shown in the upper panel of Fig. \ref{fig:pretherm_1d}. We then define the lifetime $t^*/\tau$ of the prethermal state for some $K$ as the value of $t/\tau$ for which the corresponding curve takes some reference value of $\alpha$, $\alpha=0.05$. The lifetimes $t^*/\tau$ versus $1/K$ are shown in a semi-log plot in the inset of the lower panel of Fig. \ref{fig:pretherm_1d}. The best fit to the plot points, lying almost on a straight line, is $t^*/\tau= A\exp(B/K)$ with $A=2.04$ and $B=2.08$; here $B$ depends on the reference value of $\alpha$. In accordance with this dependence of $t^*/\tau$ on $1/K$, the lower panel of Fig. \ref{fig:pretherm_1d} shows that all the curves collapse when plotted against $K{\rm log}(t/(A\tau))$.

We now explain the exponential dependence of the lifetime $t^*/\tau$ on $1/K$. Our arguments make use on the resonance theory of Arnol'd diffusion, see details in, e.g., Refs. \cite{chirikov1993theory,chirikov97arnold,chirikov79universal} by Chirikov and collaborators. According to this theory, for periodically driven many-body systems that are small perturbations of integrable systems, Arnol'd diffusion takes place along resonance ``channels" in phase space. If the system is described by $N$ conjugate pairs of action-angle variables, a {\em primary} resonance channel is a phase-space region where the following condition is satisfied [see, e.g., Eqs. (1.2), (1.5), and (1.6) in Ref. \cite{chirikov1993theory}]:
\be \label{resad}
\left |\sum_{j=1}^Nm_j\omega_j-M\Omega\right |\lesssim\sqrt{\varepsilon}\ .
\ee
Here $\{m_{j'}\}_{j'=1}^N$ are integer vectors defining the perturbation, $\omega_j$ are the frequencies of the unperturbed (integrable) system, $M$ is an arbitrary integer, $\Omega$ is the driving frequency, and $\varepsilon$ is the perturbation strength. In the case of the system (\ref{eq:H}), $\omega_j=p_j$ and the only nonzero components of a vector 
$\{m_{j'}\}_{j'=1}^N$  are $m_{j}=1$ and $m_{j+1}=-1$ for some $j$, $j=1,...,N$; also $\varepsilon =\kappa/\tau$. Equation (\ref{resad}) will then read in our case:
\be \label{resade}
\left |p_j-p_{j+1}-M\Omega\right |\lesssim\sqrt{\frac\kappa\tau}\ .
\ee

Since the prethermal state is described by the probability distribution (\ref{eq:Pstar1}) of angular momenta, the probability to satisfy the Arnol'd-diffusion condition (\ref{resade}) is given by
\begin{align}\label{PM}
P^{(M)} & = \int_{-\sqrt{\kappa}}^{+\sqrt{\kappa}} dx~ P^*\small{\left(p_j-p_{j+1} -M\Omega = x \right)} 
\\ & \approx 
2\sqrt{\frac\kappa\tau} P^*\small{\left(p_j-p_{j+1} = M\Omega \right)} 
\\
&= \left(\frac{\kappa}{\pi 	T^*\tau}\right)^{1/2} \exp\left(-\frac{M^2\Omega ^2}{4T^*}\right)\;.\nonumber
\end{align}
Here we used Eq. (\ref{eq:Pstar1}) along with the identity $P^*(p_j-p_{j+1} = x) = \int dy P^*(p_j = x+y)P^*(p_j = y) = (4\pi T^*)^{-1/2} e^{-x^2/(4T^*)}$. Using then $\Omega =2\pi /\tau$ and Eq. (\ref{Tstar}) in Eq. (\ref{PM}), we obtain
\be\label{PMK}
P^{(M)}=\left(\frac{1}{0.9381\pi}\right)^{1/2} \exp\left(-\frac{M^2\pi^2}{0.9381K}\right)\ .
\ee
If instead of primary resonances one considers high-order ones, one should replace $M$ in Eqs. (\ref{resade}) and (\ref{PMK}) by a rational number $M/L$ ($M$ and $L$ are coprime integers). Then, the time to escape from $M=0$ (corresponding to the average Hamiltonian describing the prethermal state) to resonance $M/L\neq 0$ is proportional to $1/P^{(M/L)}$. By assuming that the prethermal-state lifetime reflects the escape time to some dominant resonances $\pm M/L$ \cite{note4}, we see that this lifetime increases exponentially with $1/K$, in accordance with the numerical observations (see inset of lower panel in Fig. 5). We also note that the escape time to resonances $\pm 1/2$ is $\propto \exp\left(2.63/K\right)$, where the prefactor $2.63$ is close to the numerical one, $B=2.08$, in Fig. 5.

In conclusion, we have introduced in this paper quantitative characterizations of prethermal states in realistic periodically driven many-body systems, i.e., the paradigmatic systems (\ref{eq:H}) of many-body chaos theory, featuring an unbounded energy density. Our study substantially differs from previous ones concerning the suppression of heating in periodically driven systems with finite energy density
\cite{choudhury14stability,bukov15prethermal,abanin15exponentially,citro2015dynamical,goldman15periodically,chandran16interaction,lellouch17parametric,lellouch2018parametric,abanin15exponentially,kuwahara16floquet,mori2016rigorous,abanin17rigorous,machado17exponentially,mallayya2018prethermalization}. In the case of infinite energy density, which is generic, the systems exhibit unbounded chaotic diffusion in the asymptotic time limit, absorbing heating at an almost constant rate. Before this time limit, however, there may exist prethermal states where the kinetic energy of the system is essentially constant during a long time. We demonstrated that such a state is characterized by a GGE, where the main constant of motion is the quasi-conserved average Hamiltonian $H^*$. Also, the GGE features the temperature $T^*$ of the prethermal state. We have derived, apparently for the first time, an explicit formula, Eq. (\ref{Tstar}), for the statistical quantity $T^*$ in terms of the deterministic properties of the system ($\kappa$, $\tau$, or $K$).  

Next, we attributed the escape from the prethermal state to the encounter of many-body resonance channels. We analyzed the statistical probability of the occurrence of a resonance using the GGE description of the prethermal state. We found that this probability decreases exponentially with $\Omega^2/T^*$ or $1/K$, where $\Omega =2\pi /\tau$ is the driving frequency. For large $\Omega$, the probability is quite small, leading to prethermal states with exponentially long lifetimes, in accordance with the numerical observations. Our expression for the lifetime of the prethermal state is of pivotal importance in the field of Floquet engineering, which uses periodic drives to generate tunable couplings (see Ref.~\cite{eckardt17atomic} for an introduction). This approach often relies on the use of the average Hamiltonian as an approximation of the true many-body Floquet Hamiltonian. Hence, its validity is limited to time scales in which the average Hamiltonian is quasi-conserved. We showed that this time scale can be exponentially extended by decreasing the temperature $T^*$ in units of $\Omega^2$.

We note that the exponential scaling of the lifetime with $1/K$ is analogous to that of the bound on the Arnol'd-diffusion rate given by Nekhoroshev theorem \cite{Nek,Pos}. However, while this bound depends strongly on the number $N$ of degrees of freedom, the lifetime of the prethermal state does not depend on $N$. In fact, the effects described in this paper are not related to Nekhoroshev theorem but they rather have a {\em statistical} origin and hold for small nonintegrability strength that is much larger than that in  Nekhoroshev theorem.

\begin{acknowledgments} 
We thank D. Abanin, M. Bukov, E. Demler, A. Polkovnikov, and J. Schmiedmayer  for many useful discussions. This work is supported by the Israel Science Foundation, grant no. 1542/14.
\end{acknowledgments}


\begin{thebibliography}{38}
\bibitem[{\citenamefont{note}(2019)}]{note}
	\bibinfo{journal}{The key tool of their proofs is the Floquet theorem, which is valid for linear systems only, such as    
	quantum systems described by the Schr\"{o}dinger equation. Classical systems are described by nonlinear equations of 
	motion and are not subject to the Floquet theorem.}
\bibitem[{\citenamefont{note1}(2019)}]{note1}
	\bibinfo{journal}{Because the partition function $Z$ is a product of functions of $p_j$ and functions of $\phi_j$.}
\bibitem[{\citenamefont{note2}(2019)}]{note2}
	\bibinfo{journal}{Using Eq. (\ref{eq:Pstar1}), one can also show that the Fourier components $p_q = (1/\sqrt{N})   
	\sum_{j=1}^N \exp (i q j)(p_j-\tilde{p})$ are described by the distribution function $P^* (\{p_q \}) = Z^{-1}
	\prod_{q=1}^N \exp [-|p_q|^2/(2T^*)]$. This is in agreement with the numerical findings in Ref. 
	\cite{rajak2018stability}, where it was found that $\langle p^2_q\rangle = T^*$ does not depend on $q$.}
\bibitem[{\citenamefont{note3}(2019)}]{note3}
	\bibinfo{journal}{Because of the boundary condition $\phi_{N+1}=\phi_1$, the variables $\varphi_j$ are not independent: 
	$\varphi_N=-\sum_{j=1}^{N-1}\varphi_j$. However, for sufficiently large $N$, one may assume that the contribution of the 
	dependent variable $\varphi_N$ to Eq. (\ref{ident}) is negligible relative to the total contribution of the independent 
	variables $\varphi_j$, $j=1,...,N-1$. The agreement of the main result (\ref{Tstar}) with the numerical observations 
	confirms the validity of this assumption.}
\bibitem[{\citenamefont{note4}(2019)}]{note4}
	\bibinfo{journal}{Since different resonances generally overlap, forming the Arnol'd network     
	\cite{via,chirikov79universal}, some low-order resonances may be dominant in determining the escape rate because 
	it can be relatively easy to access from them other resonances in the Arnol'd network. We plan to study this point in 
	future works.}
\end{thebibliography}

\end{document}